\documentclass[fleqn,twoside]{article}
\usepackage{espcrc2}

\usepackage{graphicx}

\title{ Shower spectrum and detection of tau from Earth-skimming neutrinos}

\author{Ming-Huey A. Huang
\address{General Education Center, National United University \\
	1 Lien-Da, Kung-Ching Li, Miao-Li, TAIWAN. 36003, ROC}
\thanks{This study is supported by 
 National United University and 
grant NSC-92-2119-M-239-001 from National Science Council of Taiwan, 
ROC.}}

\begin{document}

\begin{abstract}
A realistic Monte-Carlo simulation code is developed to study the shower 
spectrum initiated by decay of tau leptons, which come from charged current 
interaction inside the Earth. AGN and GZK neutrino fluxes are used to simulate 
the shower spectrum. 
\end{abstract}

\maketitle

At neutrino energy above $10^{15} eV$, detecting neutrinos skimming through 
the Earth provides an unique chance to detect those high energy neutrinos and 
confirmation of  $\nu_{\mu} \rightarrow \nu_{\tau}$ oscillation. 
To evaluate the detectable event rate and optimize the detection efficiency, 
a realistic Monte-Carlo simulation code is developed.

The Earth is modeled as a spherical sphere of several layers of materials 
and density. 
 The $\nu - N$ cross-section is modeled by CTEQ-6 model~\cite{JJT}.  
 For this study, the energy loss of $\tau$ is modeled 
 by continous energy loss, i.e.,
 \[      -dE/(\rho dx) = \alpha + \beta E = \beta¡¦E, \]
where $\rho$ is density of material, $dx$ is distance traveled by $\tau$, and 
$\alpha$ and $\beta$ is the coefficients of energy loss from all 
processes~\cite{JJT}.
The $\tau$ survival probability $P$ is modeled by 
\[ dP/P = -dx/(\kappa E) = dE/(\kappa \beta¡¦\rho E^2),\] 
where $\kappa E$ is decay length of $\tau$ at energy E in PeV and 
$\kappa$ = 49.9 m/PeV. 

$\tau$ decay is simulated with TAUOLA package with fully polarized $\tau$s. 
When $\tau$s decay inside the Earth, the $\nu_{\tau}$s are re-propagated 
inside the Earth and could generate new $\tau$s again. 
When a $\tau$ decays outside the Earth, it initiate air shower, except when 
$\tau$ decay to muon. The shower energy is sum of the energy from electrons 
and hadrons. The shower simulation is done by a modified version of 
CORSIKA~\cite{MAH03}. 

To compare with previous results from analytical method~\cite{JJT}, 
$\nu_{\tau}$s are injected to a fix width of 100km of rock. 
Shown in the Fig. 1, the $\tau$ spectrum from both methods agrees quite well. 
This test conclude that the algorithm used in this Monte-Carlo simulation is 
correct. We are updating energy loss with Stochastic processes and the 
complete study will be published soon.
\vspace{-1pc}
\begin{figure}[htb]
\includegraphics[width=8.5pc]{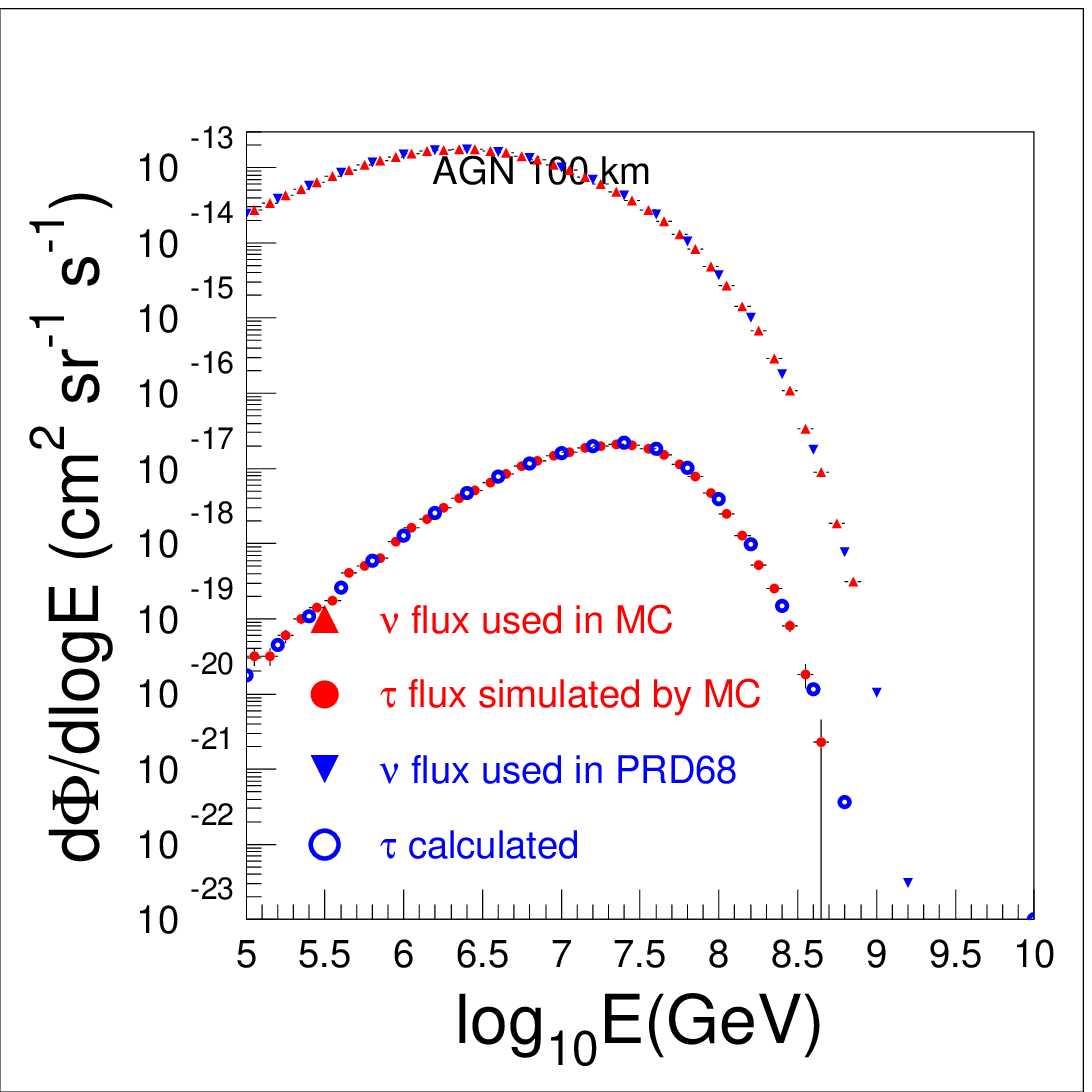}
\includegraphics[width=8.5pc]{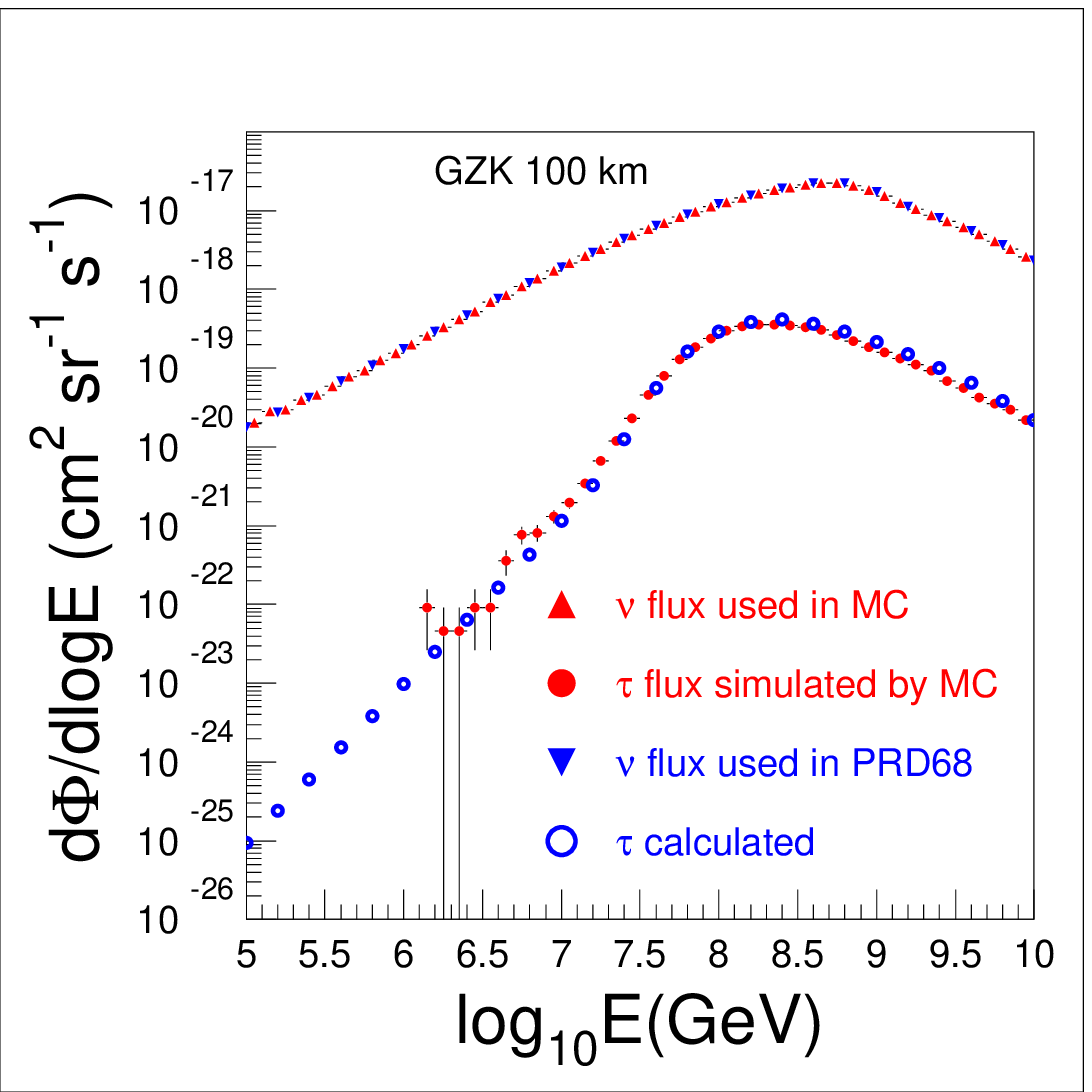}
\vspace{-2pc}
\caption{Spectrum of $\tau$ from  $\nu_{\tau}$ pass through 100 km of 
rock from AGN (left figure)~\cite{AGNnu} and 
GZK (right figure)~\cite{GZKnu}. The blue markers come from analytical 
calculation~\cite{JJT}, while the red markers come from Monte-Carlo simulation 
in this study. 
}
\label{fig:spectrum}
\end{figure}

\end{document}